\documentclass[aps,pra,reprint,groupedaddress]{revtex4-2}

\usepackage{amssymb,amsmath}
\usepackage{graphicx}
\usepackage{xcolor}
\usepackage{bm}% bold math
\usepackage{dcolumn}
\usepackage[pdftex,breaklinks,colorlinks,citecolor=blue,linkcolor=blue,urlcolor=blue]{hyperref}

\usepackage{mathrsfs} % For mathscr
\begin{document}

\title{Arrangement and chemical complexity in an $N$-body quantum system}

\author{Bo Gao}
\email[Email: ]{bo.gao@utoledo.edu}
\homepage[Homepage: ]{http://bgaowww.physics.utoledo.edu}
%\thanks{}
\affiliation{Department of Physics and Astronomy,
	University of Toledo, Mailstop 111, 
	Toledo, Ohio 43606,
	USA}

\date{November 13, 2022}
%\date{\today}

\begin{abstract}

We introduce and discuss the concept of \textit{arrangement}, traditionally found in the context of chemical reactions and few-body rearrangement collisions, in the general context of an $N$-body quantum system. We show that the ability for particles to attract and bind is a key source of complexity of an $N$-body quantum system, and \textit{arrangement} is an emergent concept necessitated by the description of this complexity. For an $N$-body system made of particles of which some or all can attract and bind, the concept of arrangement is not only necessary for a full characterization of its quantum states and processes, but it also provides a basis for the understanding and the description of the structure of its energy spectrum. We also discuss and formulate multiparticle separability in an $N$-body system, the mathematical foundation that underlies the arrangement concept in an $N$-body theory.

\end{abstract}

%\keywords{}

\maketitle

\section{Introduction}

In a recent work \cite{Gao22a}, we reconstructed the foundation for an $N$-body quantum theory using concepts of rigidity, 2-particle separability, and 2-particle cusp conditions. The theory established, in particular, the boundary conditions around 2-particle coalescence points that a physical solution should satisfy but which are not contained in the Schr\"odinger equation. We mentioned that while the formulation of Ref.~\cite{Gao22a} should be sufficient for systems of particles that only repel each other, we needed at least one more piece of the foundation, built around the concept of \textit{arrangement}, for systems of particles of which some or all can attract and bind. This work is intended to be that missing piece: a piece necessary to complete the description of a general $N$-body state and/or process.

We introduce here the concept of \textit{arrangement} as an emergent concept in transitioning from a 2-body to a $(N\ge 3)$-body system, similar to the emergence of the length scale $r_{\rho}$ of Ref.~\cite{Gao22a}, but unique to $N$-body quantum systems with 2 or more of their constituent particles able to attract and bind. For simplicity, we will often call such systems $N$-body systems with binding, while calling systems in which no configuration of particles can bind $N$-body systems without binding. We will show that $N$-body systems with binding are qualitatively more complex than systems without binding, and the concept of arrangement is an emergent concept necessitated by the description of this complexity. For systems with binding, the arrangement concept is \textit{necessary}, not optional, to fully characterize their quantum states, and is required to properly count states and evaluate entropy. It is also a key concept describing the complexity of such systems, to be called the chemical complexity, that arises from binding and is beyond the total number of states and beyond that of nonbinding systems. We will also show that the arrangement can be defined to be a general concept for all quantum systems as it can be naturally extrapolated to cover cases, such as 2-body systems, where it is \textit{not} necessary.

The paper is organized as follows. In Sec.~\ref{sec:complexity}, we provide motivation for the concept of arrangement through a discussion of the complexity of an $N$-body quantum system, specifically the qualitative differences in complexity between systems with and without binding. In Sec.~\ref{sec:arrDef}, we define arrangement for an arbitrary $N$-body quantum system and suggest a general notation and an optional numbering. In Sec.~\ref{sec:arrInTheory}, we take a closer look at the arrangement concept from the perspective of an $N$-body quantum theory, including a discussion of multiparticle separability which is the mathematical foundation of the arrangement concept. We also introduce in Sec.~\ref{sec:arrInTheory} the total number of arrangements as a measure of the chemical complexity that arises from binding. The related discussion further helps to illustrate the differences in complexity between systems of distinguishable particles and systems of identical particles. We conclude in Sec.~\ref{sec:conc} with a summary, and a brief discussion of the expanding roles of the $N$-body quantum theory in other areas of sciences.

\section{Complexity of an $N$-body quantum system}
\label{sec:complexity}

\begin{figure}
\includegraphics[width=\columnwidth]{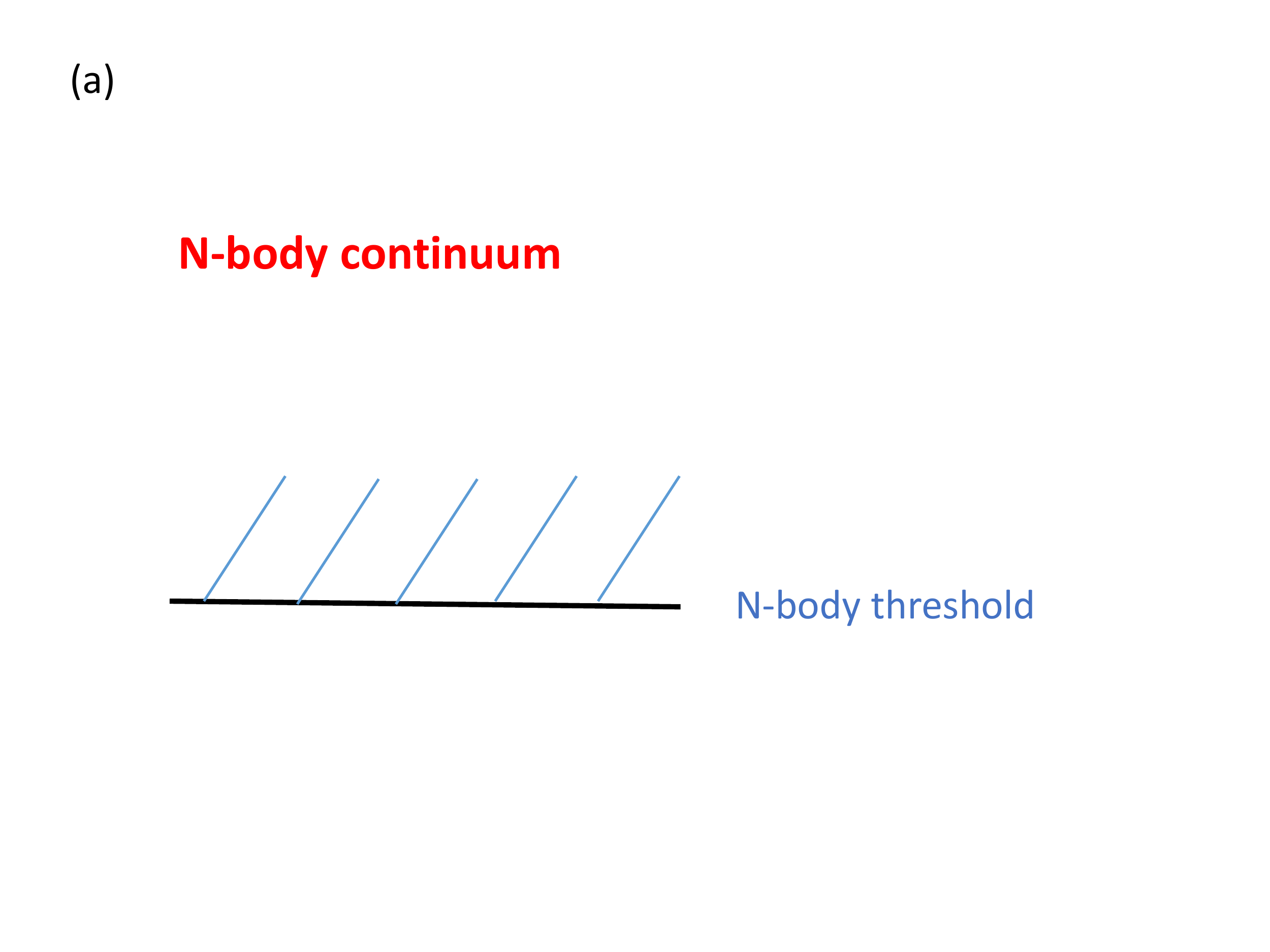}
\includegraphics[width=\columnwidth]{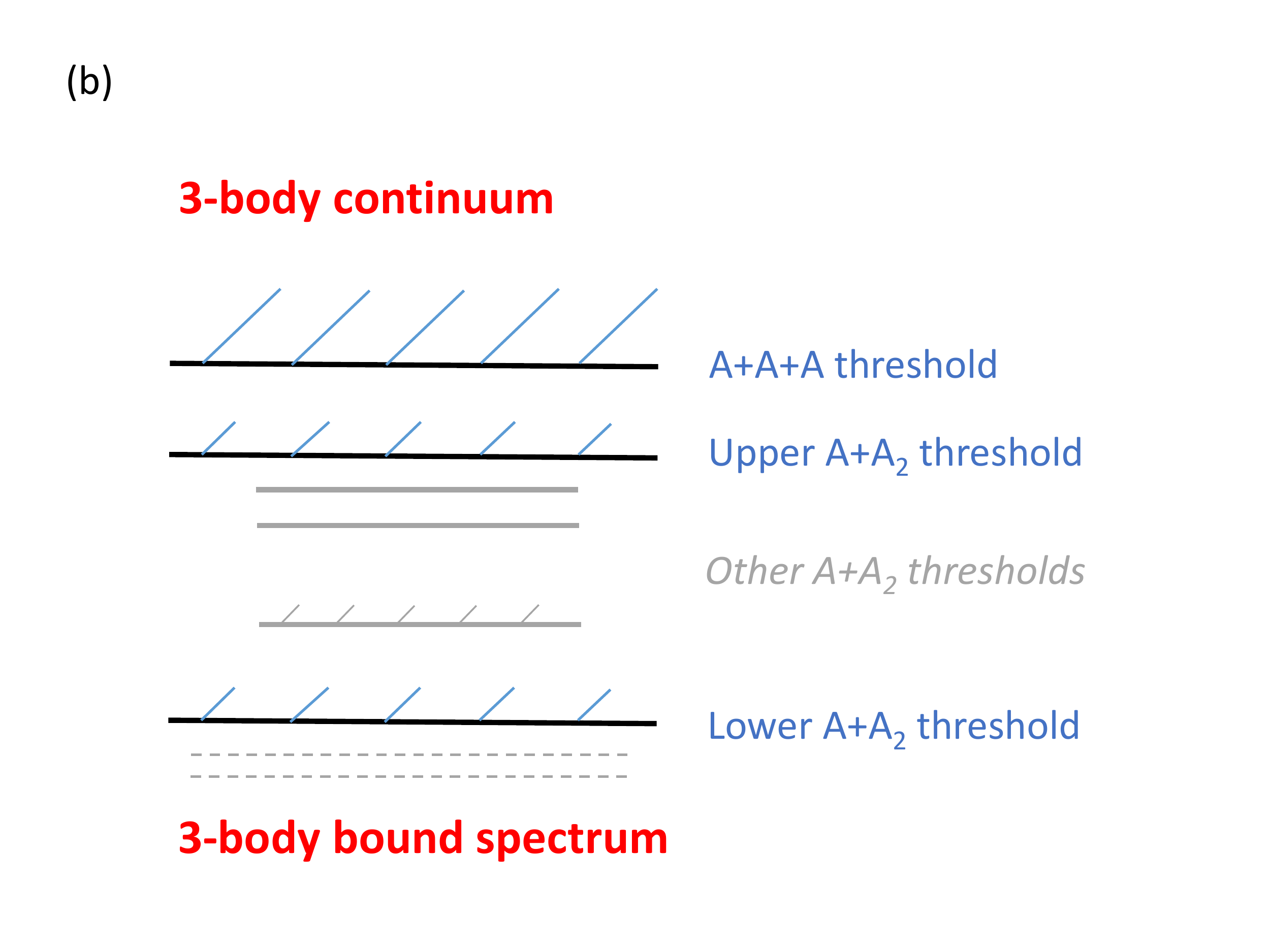}
\caption{(Color online) (a) Energy spectrum for $N$ free particles or particles that only repel each other. (b) Energy spectrum for three identical particles that can bind.
\label{fig:Nb3bCmp}}
\end{figure}

In this section, we discuss briefly the role of interaction in complexity, more specifically the fundamental differences in complexity between quantum systems with and without binding. It serves both as a motivation and an introduction to the concept of arrangement.

Consider a quantum system of $N$ particles. It is well recognized that such a system can quickly become ``complex'' beyond $N=2$. This growth in complexity is often associated with or explained as the exponential growth of the dimension of the Hilbert space. Specifically, if we need $b$ basis functions to characterize a single particle, we will need $b^N=e^{N\ln b}$ basis functions to characterize $N$ such particles, a number that grows exponentially with $N$. This measure of ``complexity'' in terms of the dimension of the Hilbert space, or equivalently the total number of states, is useful, but far from complete. It can in fact be misleading especially if we want the word ``complexity'' to refer more specifically to the complexity of the behaviors of a quantum system. 

The example of $N$ free (meaning non-interacting here) particles is sufficient to illustrate this point. The thermodynamics of such systems are known analytically \cite{LandauSTM1,Huang87}, and they are not complex. This remains true even if the particles themselves have a spin and/or other internal states \cite{LandauSTM1,Huang87}. It leads to the first point to note about the complexity of an $N$-body quantum system: If particles do not interact, no complex behavior emerges even for large $N$. Keeping in mind that interaction does not change the total number of states, i.e. the dimension of the Hilbert space, we can also conclude that the complex behavior of a quantum system has little to do with the total number of states. The simplicity of $N$ free particles further manifests in the simplicity of its energy spectrum, as illustrated in Fig.~\ref{fig:Nb3bCmp}(a).

Next consider an $N$-body system in which the particles do interact, but can only repel each other. The behavior of such a system can again be expected to remain simple, as reflected in its energy spectrum having a single $N$-body continuum as illustrated in Fig.~\ref{fig:Nb3bCmp}(a). The structure of the continuum is furthermore expected to remain simple if the repulsion is everywhere more repulsive than $+1/r^2$, for which not even shape resonances are to be expected \cite{Gao17a}.

Now consider a system of particles in which some or all of the particles can attract and bind. For simplicity, consider the simplest case of 3 identical particles, such as 3 identical atoms that can attract and bind in all configurations. Figure~\ref{fig:Nb3bCmp}(b) gives an illustration of its energy spectrum. It is astonishing how this spectrum, basically the simplest among all $N$-body systems with binding, compares in complexity to those without binding. Below the lowest $A+A_2$ threshold, corresponding to $A_2$ in its lowest-energy bound state (the ground state), we have a discrete set of 3-body bound spectrum, corresponding to states in which all 3 particles are bound. Above the lowest $A+A_2$ threshold, we have a continuum of states in which the 3 particles can always break into an $A$ and an $A_2$ with extra energy shared as their kinetic energies, until above the $A+A+A$ threshold where the particles have an extra option of breaking into 3 free particles. Between the lowest $A+A_2$ threshold and the $A+A+A$ threshold, the continuum is characterized by a set of thresholds corresponding to $A_2$ in its progressive more excited states, until the uppermost $A+A_2$ threshold corresponding to $A_2$ in its most loosely bound state. And below each one of these thresholds, we have generally a set of Feshbach resonances that further complicates the structure of the continuum.

In contrast to the simplicity of $N$ free particles or $N$ particles without binding, even a simple case of 3 particles with binding can already be so complex to be beyond the reach of our current brute-force computational methods (see, e.g., Refs.~\cite{Quemener05,Mayle12,Mayle13,Gregory19}). This simple example illustrates the fundamental difference in complexity between $N$-body quantum systems with and without binding. This fundamental difference is characterized by the concept of \textit{arrangement} that we have utilized here in a simple 3-body example and will be generalized to an arbitrary $N$-body system in the following sections.

\section{Concept of arrangement}
\label{sec:arrDef}

\subsection{The example of a 3-body system}

We return first to a more complete discussion of the 3-body example, starting with the case of 3 distinguishable particles $A$, $B$, and $C$, that can bind in all configurations. It is an example where one already has exposure to the concept of arrangement either through chemical reactions \cite{Levine05} or through rearrangement collisions \cite{Taylor06}.

A self-evident and familiar representation of chemical reactions and/or rearrangement collisions for such 3 particles is given by the chemical equation
\begin{subequations}
\begin{align}
A+(BC) &\to B+(AC) \;,\\
	&\to C+(AB) \;,\\
	&\to A+B+C \;,
\end{align}
\label{eq:3PchemRex}
\end{subequations}
in which we have used the parentheses to more clearly identify a bound group (a ``molecule''). This self-evident representation of 3-particle states incorporates the fact that the 3 particles can be bound differently, and given sufficient energy, different configurations of binding can convert into each other.

This different configuration of binding, or different grouping of particles into bound groups, is characterized by the concept of arrangement. In terms of notation, the plus signs, while convenient in chemical equations to separate different, relatively unbound groups, are more conveniently and generally more concisely represented by parentheses in a text. Thus the arrangement of $A+B+C$ in an equation is to be represented by $(A)(B)(C)$ in a text, and similarly for other arrangements.

Equation~(\ref{eq:3PchemRex}) contains, in our definition, 4 out of all 5 arrangements for 3 distinguishable particles that can bind in all configurations. The 5 arrangements are, in text representation, $(A,B,C)$, $(A)(B,C)$, $(B)(A,C)$, $(C)(A,B)$, and $(A)(B)(C)$, of which the last 4 have their equation representations in Eq.~(\ref{eq:3PchemRex}). The arrangement $(A,B,C)$, which can be represented in an equation as $(ABC)$ is the arrangement in which all 3 particles are bound. Its absence from Eq.~(\ref{eq:3PchemRex}) is a reflection that for energetic reasons, a true 3-body bound state would not break up by itself without another particle, which can be a photon, supplying the necessary energy. And similarly, any of the other 4 arrangements cannot become a true 3-body bound state without another particle, e.g. a photon, taking away energy. This absence of conversion has often excluded the configuration of all-bound as one of the arrangements. As a general concept for an $N$-body system, however, it is better, both conceptually and mathematically, to include the configuration of all-bound, when it exists, as one of the arrangements in order to cover the entire energy spectrum of an $N$-body system [see, e.g., Fig.~\ref{fig:Nb3bCmp}(b)].  With this inclusion, photon-assisted processes such as photoionization/dissociation/detachment
\begin{subequations}
\begin{align}
(ABC)+\gamma &\to A+(BC) \;,\\
	&\to B+(AC) \;,\\
	&\to C+(AB) \;,\\
	&\to A+B+C \;,
\end{align}
%\label{eq:3PchemRex}
\end{subequations}
are all regarded as rearrangement processes, specifically as photoinduced rearrangement processes.

A simple extrapolation of this example serves to illustrate that the number of arrangements depends on whether the particles are identical or distinguishable. If the 3 particles that we have considered are identical, the arrangement $(A,B,C)$ in the distinguishable case becomes $(A,A,A)$, more conveniently represented as $(A_3)$. The arrangements $(A)(B,C)$, $(B)(A,C)$, and $(C)(A,B)$ all become the same, a single arrangement of $(A)(A,A)$, more conveniently represented as $(A)(A_2)$. The arrangement $(A)(B)(C)$ becomes $(A)(A)(A)$, more conveniently represented as $(A)_3$. Thus the 5 arrangements for 3 distinguishable particles have become 3 arrangements for 3 identical particles. This difference in the number of arrangements for identical and distinguishable particles is much more dramatic for a larger $N$, as will be further discussed and illustrated in Sec.~\ref{sec:chemComplexity}. 

\subsection{Definition and notation for an $N$-body system}

We define \textit{arrangement} for an $N$-body system as a property of its states that describes the grouping of its constituent particles into bound groups and (unbounded) free particles, with the latter being regarded as a special ``bound group'' with only a single particle. The particles within the same group, if more than 1, are bound to each other, while different groups are unbound relative to each other. In other words, different groups can move to the infinite separation where they are free from each other with a positive relative kinetic energy. An arrangement can be called more precisely a \textit{binding-arrangement} if there is any possibility of confusion with other terminologies.

An arrangement can be denoted using a notation that is consistent both in representing specific real-world systems and in an abstraction of an $N$-body theory. In an abstract notation, we would use symbols such as $A, B, C$ to denote a generic particle. Different groups are separated by parentheses with repetition contracted into a subscript. For instance, the arrangement of $N$ free particles, none bound, can be represented as $(A)_N$. This is the typical arrangement associated with an atomic gas. Its thermodynamic limit of $N\to\infty$ can be consistently represented as $(A)_{\infty}$. The symbol $(A_2)_{N/2}$ would represent the (gaseous) arrangement of $N/2$ dimers. And $(A_X)(A)_{N-X}$ represents an arrangement in which $X$ particles $A$ are in a bound group (a cluster or a droplet or a crystal), while there are $N-X$ other unbounded particles $A$.

In representing a specific physical system, in what we call a chemical notation, we would use chemical symbols to represent the particles. For instance, $(Rb)_\infty$ would represent the gaseous arrangement of Rb atoms in the thermodynamic limit, with none bound. The symbol $(Rb_2)(Rb)_{\infty}$ would represent an arrangement of a single Rb$_2$ molecule in a gas of Rb atoms. 

\subsection{Structure of the energy spectrum and an optional numbering}

The concept of arrangement helps to describe the complex energy spectrum of an $N$-body quantum system with binding. By including the arrangement of all-bound, when it exists, the concept provides a description of the structure of the $N$-body energy spectrum that covers the entire spectrum. The arrangement of all-bound covers the lowest energy region corresponding to the (in-principle-discrete) bound spectrum for all $N$ particles. All other arrangements describe continuum states with one or more of the particles and/or bound groups with sufficient energy to break free. Among all these continuum arrangements, the arrangement of all-free has a single highest threshold. All other continuum arrangements, with at least one or more bound groups, have generally a set of thresholds, with the lowest threshold corresponding to each bound group being in its lowest energy state, and the highest threshold corresponding to each bound group being in its highest energy state (before breaking up into a different arrangement with one more free particle). The earlier example in Sec.~\ref{sec:complexity} for 3 identical particles gave one of the simplest illustrations of this general qualitative picture for an $N$-body system with binding.
 
This qualitative characterization of an energy spectrum using arrangements can be made more quantitative by optionally assigning each arrangement a non-negative integer $g$, starting either at 0 or 1, that reflects the energy ordering of the arrangements. More specifically, the arrangement of all-bound, if exists, is numbered as $g=0$. All other arrangements are numbered, starting from 1, according to the energy ordering of their lowest threshold. With such a numbering system and letting $\mathcal{T}_{g}$ to denote the lowest threshold for arrangement $g$, the definition of $g$ is such that for energies between $\mathcal{T}_{g}$ and $\mathcal{T}_{g+1}$, there are $g$ arrangements open to rearrangement collisions, and the continuum states in this region are $g$-fold degenerate in arrangements.

In this numbering scheme, the arrangement of all-bound, when it exists, is always numbered as $g=0$. The arrangement of all-free always has the greatest $g$ since it has the highest energy threshold.  Let $\mathscr{M}(N)$ be the total number of arrangements for an $N$-body quantum system (see Sec.~\ref{sec:chemComplexity} for more discussion), the arrangement of all-free is numbered either as $g = \mathscr{M}(N)-1$ if the arrangement of all-bound exists, or as $g = \mathscr{M}(N)$ if the arrangement of all-bound does not exist. For other continuum arrangements in between all-bound and all-free, there are cases where their energy ordering, thus their numbering, is obvious. There are also other cases where the energy ordering is not obvious and its determination is part of understanding the $N$-body system.

For the example in Sec.~\ref{sec:complexity} of 3 identical particles able to bind in all configurations, 
the arrangement of all-bound, $(A_3)$, has $g=0$. The arrangement of $(A)(A_2)$ has $g=1$. The arrangement of all-free, $(A)_3$, has $g=2$. This is one of the examples where the energy ordering of the arrangements is obvious.

As another example, a neutral atom of atomic number $Z$, when regarded as an $N$-body quantum system made of 1 nucleus and $Z$ electrons, has $Z+1$ arrangements: $(A)$ ($g=0$), $(A^+)(e^-)$ ($g=1$),\dots, $(A^{Z+})(e^-)_Z$ ($g=Z$). This is another example where the energy ordering of the arrangements is obvious.

A 3-body system of one atom $A$ and two electrons has 2 arrangements: $(A^-)(e^-)$, corresponding to one atomic negative ion $A^-$ and one free electron, labeled as $g=1$, and $(A)(e^-)_2$, corresponding to one atom and two free electrons, labeled as $g=2$. There is no arrangement corresponding to the all-bound configuration of $A^{2-}$, as there is no known atom that can bind 2 electrons. This is an example of an $N$-body system with binding but with no arrangement of all-bound.

The earlier example of 3 distinguishable particles able to bind in all configurations is an example for which the ordering of the in-between continuum arrangements requires more detailed knowledge as a part of understanding the $N$-body system. Out of its total 5 arrangements, the arrangement of all-bound, $(A,B,C)$ has $g=0$. The arrangement of all-free, $(A)(B)(C)$ has $g=4$. The ordering of the other 3 in-between continuum arrangements: $(A)(B,C)$, $(B)(A,C)$, and $(C)(A,B)$, requires more knowledge about which of the pairs, $(B,C)$, $(A,C)$, and $(A,B)$, has the more tightly bound ground state, for which the corresponding arrangement would have a smaller $g$.

\section{Arrangement in $N$-body theory}
\label{sec:arrInTheory}

\subsection{An emergent concept extrapolated to cover all cases}

We already know from existing theories that the concept of arrangement is not necessary for 1- and 2-body systems, nor is it necessary for systems of particles that only repel and are not capable of binding. In this sense, the concept of arrangement is best understood as an emergent property in going from a 2- to an $(N\ge 3)$-body system, for $N$-body systems with binding. For such systems, the concept of arrangement is \textit{necessary} for a full characterization of their continuum states which are $g$-fold degenerate in arrangements for energies between $\mathcal{T}_{g}$ and $\mathcal{T}_{g+1}$, and the emergence of this concept characterizes the most important qualitative change from an $N$-body system without binding. We note that the necessity of the concept is not only for the description of rearrangement collisions but also in other fundamental aspects such as the counting of states and therefore the proper evaluation of entropy.

With this understanding in place, it is also useful to note that the concept of arrangement can be easily extrapolated to cover all $N$-body systems. This allows us to have a single $N$-body theory for all cases instead of different theories for different cases. In such an extrapolation, the 1-body case and the $N$-body case with only repulsive interactions both have only a single arrangement, represented, e.g., by $(A)$ ($g=1$) and $(A)_{n_A}(B)_{n_B}\cdots$ ($g=1$), respectively. The case of a 2-body system capable of binding has two arrangements, $(A,B)$ ($g=0$) and $(A)(B)$ ($g=1$), though the arrangement concept is not absolutely necessary in this case since all continuum states belong to a single arrangement.

\subsection{$N_{sb}$-particle separability}

Behind the concept of arrangement, there is the hidden $N_{sb}$-particle separability ($N_{sb}\ge 2$) that is its mathematical basis and representation. We have derived and discussed the 2-body separability in detail in the earlier paper \cite{Gao22a} and briefly mentioned there the $N_{sb}$-particle separability for $N_{sb}>2$. We expand here on this topic and discuss its relation to the concept of arrangement.

Consider an $N$-body quantum system with $N>3$ interacting via pairwise potentials. It is described by a Hamiltonian
\begin{equation}
H^{(N)} = \sum_{i=1}^N -\frac{\hbar^2}{2m_i}\nabla_i^2 + \sum_{i<j=1}^N \hat{v}_{ij}(r_{ij})\;,
\label{eq:nbH}
\end{equation}
and a corresponding Schr\"odinger equation at an energy $E$
\begin{equation}
H^{(N)}\Psi^{(N)}_{E\gamma} = E\Psi^{(N)}_{E\gamma} \;.
\end{equation}
Here $\gamma$ represents all quantum numbers required to uniquely specify an $N$-body state. $\hat{v}_{ij}(r_{ij})$ is the interaction potential between particles $i$ and $j$. It is generally an operator if more-than-one internal states, including spin states, are involved in the energy range of interest. For simplicity, but without loss of generality for our purposes, we assume $\hat{v}_{ij}(r_{ij}) = v_{ij}(r_{ij})$ to be a scalar central potential. The cases of spin-dependent, multichannel, and/or anisotropic potentials \cite{Gao20b} will be addressed in future studies.

For any chosen $N_{sb}$ ($2<N_{sb}<N$) out of $N$ particles, which may, or may not, be identical, the Hamiltonian can always be regrouped, using the center-of-mass frame of the chosen $N_{sb}$ particles, as 
\begin{equation}
H^{(N)} = h_{N_{sb}}+\widetilde{H}^{(N)}_{N_{sb}} \;.
\end{equation}
Here
\begin{equation}
h_{N_{sb}} := H^{(N_{sb})} - \left(-\frac{\hbar^2}{2 M_{N_{sb}}}\nabla_c^2\right) \;,
\end{equation}
describes the \textit{relative} motion of the $N_{sb}$-particles subsystem in its center-of-mass frame with its center-of-mass $\bm{c}$ and its total mass $M_{N_{sb}}$ defined by 
\begin{subequations}
\begin{align}
& \bm{c} := \frac{1}{M_{N_{sb}}} \sum_{i=1}^{N_{sb}} m_i\bm{r}_i \\
& M_{N_{sb}} := \sum_{i=1}^{N_{sb}} m_i \;.
\end{align}
\end{subequations}
The $\widetilde{H}^{(N)}_{N_{sb}}$ is the Hamiltonian for all other degrees of freedom that include the other $N-N_{sb}$ particles and the center-of-mass motion of the $N_{sb}$-particle subsystem. It includes also the interaction between the $N_{sb}$-particle subsystem and the other particles. Specifically,
\begin{equation}
\widetilde{H}^{(N)}_{N_{sb}} := H^{(N-N_{sb})}+\left(-\frac{\hbar^2}{2 M_{N_{sb}}}\nabla_c^2\right)+\widetilde{V}^{(N)}_{N_{sb}} \;,
\end{equation}
where $H^{(N-N_{sb})}$ is the Hamiltonian describing the $(N-N_{sb})$-particle subsystem, $-\tfrac{\hbar^2}{2 M_{N_{sb}}}\nabla_c^2$ is the kinetic energy of the center-of-mass of the $N_{sb}$-particle subsystem, and
\begin{equation}
\widetilde{V}^{(N)}_{N_{sb}} := \sum_{j=N_{sb}+1}^N \sum_{i=1}^{N_{sb}} v_{ij}(r_{ij}) \;,
\end{equation}
is the interaction between the $N_{sb}$-particle subsystem and the rest, the $(N-N_{sb})$-particle subsystem.

Let $R$ be the hyperradius \cite{Delves60,Avery18,Greene17} for the $N_{sb}$-particle subsystem. It is a measure of the overall size of the subsystem, which can be understood through its relation to the distances of the particles from their center-of-mass $\bm{c}$ \cite{Delves60}, 
\begin{equation}
\mu R^2 = \sum_{i=1}^{N_{sb}} m_i r_{ci}^2 \;,
\end{equation}
where $r_{ci}:= |\bm{r}_i-\bm{c}|$ is the distance from particle $i$ to $\bm{c}$, and
\begin{equation}
\mu = \left(\frac{1}{M_{N_{sb}}}\prod_{i=1}^{N_{sb}} m_i \right)^{1/(N_{sb}-1)} \;,
\end{equation}
is the $N_{sb}$-body reduced mass. For any fixed $R$, $r_{ci}$ is confined by
\[
r_{ci} < \sqrt{\mu/m_i} \: R \;,
\]
for all $i = 1,\dots,N_{sb}$.

With this understanding and noting that $r_{ij}=|\bm{r}_j-\bm{r}_i|=|\bm{r}_{cj}-\bm{r}_{ci}|$, where $\bm{r}_{cj}:=\bm{r}_j-\bm{c}$, it is clear that for basically all physical potentials \cite{Gao22a},
\begin{equation}
\widetilde{V}^{(N)}_{N_{sb}} \overset{R\to 0}{\sim} {V}^{(N)}_{N_{sb}} 
	:= \sum_{j=N_{sb}+1}^N \sum_{i=1}^{N_{sb}} v_{ij}(r_{cj}) \;,
\label{eq:NbVsep}
\end{equation}
meaning that $\widetilde{V}^{(N)}_{N_{sb}}$ becomes, in the limit of $R\to 0$, independent of relative coordinates of the $N_{sb}$-particle subsystem. Consequently, we have
\begin{equation}
\widetilde{H}^{(N)}_{N_{sb}} \overset{R\to 0}{\sim} {H}^{(N)}_{N_{sb}} 
	:= H^{(N-2)}+\left(-\frac{\hbar^2}{2 M_{N_{sb}}}\nabla_c^2\right)+{V}^{(N)}_{N_{sb}} \;,
\end{equation}
becoming independent of relative coordinates of the $N_{sb}$-particle subsystem, and
\begin{equation}
H^{(N)} \overset{R\to 0}{\sim} H^{(N)}_{sp} := h_{N_{sb}}+{H}^{(N)}_{N_{sb}} \;,
\label{eq:NbHsep}
\end{equation}
becomes two parts that are independent of each other. We are using here the subscript of $sp$ to represent the results in the separable limit. 

The separability of the $N_{sb}$-particle relative motion, described by $h_{N_{sb}}$, from the other degrees of freedom, implies that in the limit of $R\to 0$, the $N$-body wave function can always be written as a linear superposition of product states, specifically,
\begin{equation}
\Psi^{(N)}_{E\gamma} \overset{R\to 0}{\sim} 
	\sum_{\lambda\alpha'\gamma'} a_{\lambda\alpha'\gamma'}
	\psi^{(N_{sb})}_{\lambda\alpha'}
	\Psi^{sp}_{(E-\lambda)\gamma'} \;.
\label{eq:NbWfnSep}
\end{equation}
Here $\psi^{(N_{sb})}_{\lambda\alpha}$ is the wave function for the $N_{sb}$-particle relative motion at an $N_{sb}$-particle relative energy $\lambda$,
\begin{equation}
h_{N_{sb}}\psi^{(N_{sb})}_{\lambda\alpha} 
	=  \lambda\psi^{(N_{sb})}_{\lambda\alpha} \;.
\end{equation}
$\Psi^{sp}_{(E-\lambda)\gamma'}$ is an eigenfunction of ${H}^{(N)}_{N_{sb}}$ at energy $E-\lambda$, for the other $N-N_{sb}$ particles plus the  center-of-mass motion of the $N_{sb}$-particle subsystem, defined by
\begin{equation}
{H}^{(N)}_{N_{sb}}\Psi^{sp}_{(E-\lambda)\gamma'}
	= (E-\lambda)\Psi^{sp}_{(E-\lambda)\gamma'} \;.
\end{equation}
The sum over direct products, with $a_{\lambda\alpha'\gamma'}$ being the constant coefficients of superposition, is constrained by symmetries and conservation laws, with details determined by the degree to which the $N_{sb}$-particle states are mixed by interaction with other $N-N_{sb}$ particles \textit{outside} of the region of separability.  In particular, the sum over the $N_{sb}$-particle relative energy $\lambda$, which generally includes both a sum over a discrete spectrum and an integration over a continuum spectrum for an $N_{sb}$-particle subsystem that can bind, describes a mixing of $N_{sb}$-particle wave functions of different energies. At the simplest qualitative level, this mixing can be characterized by a mean $\bar{\lambda}$, and a width $\Delta\lambda$ which gives a measure of the degree of mixing. We call Eq.~(\ref{eq:NbWfnSep}), a rigorous result on an $N$-body wave function, the $N_{sb}$-particle separability condition of an $(N>2)$-body wave function. It is a generalization of the 2-particle separability condition of Ref.~\cite{Gao22a}.

This derivation shows that an $N_{sb}$-particle relative motion in an $N$-body system is always separable, for sufficiently small $R$, from the other degrees of freedoms for all physical potentials that are continuous in $(0,\infty)$ and once or twice differentiable \cite{Gao22a}. 
Similar to the discussion of the 2-particle separability in Ref.~\cite{Gao22a}, the average value of $r_{cj} = |\bm{r}_j-\bm{c}|$ in Eq.~(\ref{eq:NbVsep}), namely the mean separation of the center-of-mass of $N_{sb}$-particle subsystem from other particles, defines an emerging length scale $r_{\rho}$. For all $N$-body systems with electromagnetic type interactions, which include all atomic and molecular interactions, the separable limit of $R\to 0$ is more precisely $R/r_{\rho}\to 0$, or characterized as an asymptotic region as the region of $R\ll r_{\rho}$ \cite{Gao22a}. Since systems of electromagnetic type are our primary focus, we rewrite, explicitly for such systems, the separable limit and the wave function separability condition more precisely as
\begin{equation}
H^{(N)} \overset{R\ll r_{\rho}}{\sim} h_{N_{sb}}+{H}^{(N)}_{N_{sb}} \;,
\label{eq:NbHsepEM}
\end{equation}
and
\begin{equation}
\Psi^{(N)}_{E\gamma} \overset{R\ll r_{\rho}}{\sim} 
	\sum_{\lambda\alpha'\gamma'} a_{\lambda\alpha'\gamma'}
	\psi^{(N_{sb})}_{\lambda\alpha'}
	\Psi^{sp}_{(E-\lambda)\gamma'} \;.
\label{eq:NbWfnSepEM}
\end{equation}

The $N_{sb}$-particle separability is a general characteristic of an $(N>N_{sb})$-body system, regardless of its binding characteristics. Equation~(\ref{eq:NbWfnSepEM}) implies the states of an $N_{sb}$-particle subsystem is generally mixed by its interaction with other particles. Importantly, it also implies, in cases where the $N_{sb}$-particle subsystem can bind, that for sufficiently low densities of other particles, leading to sufficiently large $r_\rho$, there exist $N$-body states describing an $N_{sb}$-particle bound system in the presence of other $N-N_{sb}$ particles: 
\begin{equation}
\Psi^{(N)}_{E\gamma} \approx \Psi^{(N,N_{sb})}_{E\gamma, \lambda_n\alpha}
	\overset{R\ll r_{\rho}}{\sim}
	\psi^{(N_{sb})}_{\lambda_n\alpha}\Psi^{sp}_{(E-\lambda_n)\gamma'} \;,
%\label{eq:NbWfnSepEM}
\end{equation}
in which the mixing due to the interactions with other particles is negligible or can be taken into account with a proper choice of $N_{sb}$-particle quantum numbers $\alpha$. This is the mathematical foundation for an $N_{sb}$-particle bound state in an $N$-body quantum system, thus also the foundation for the concept of arrangement. It is in cases where the bound groups in an arrangement are not too strongly mixed by their interactions with other groups that the concept of arrangement is the most useful. These are also cases where one can in principle prepare the bound groups in their own eigenstates and carry out experiments that can directly reflect the dynamics of the $N$-body system under investigation.

\subsection{Chemical complexity}
\label{sec:chemComplexity}

Being part of a qualitative characterization of an $N$-body energy spectrum, the number of arrangements, $\mathscr{M}(N)$, provides a measure of the complexity of an $N$-body system due to binding. We call this complexity associated with binding the chemical complexity, as it is, in essence, what distinguishes chemistry from $N$-body systems without binding.

Depending on the composition of an $N$-body system, specifically on how many of the particles can bind and the configurations in which they can bind, and on whether the particles are distinguishable or identical, the number of arrangements and thus the complexity of an $N$-body spectrum, can be \textit{vastly} different. While this number is not difficult to find for individual cases, it is nontrivial in general. Through considerations of some extreme cases, we provide here both rigorous constraints and a better qualitative understanding of the number of arrangements and thus the chemical complexity.

First, we note that the minimum number of arrangements is 1, describing the cases where none of the particles can bind in any configuration, such as a system of $N$ electrons or a system of $N$ protons or positive ions. Note that the number 1 is independent of $N$, a reflection that a large $N$ does not automatically imply a complex system.

\begin{table}
\caption{The number of arrangements for $N$ particles, able to bind in all configurations, lies between two extreme cases with the lower limit $p(N)$ corresponding to the case of all particles being identical and the upper limit, $B(N)$, corresponding to all particles being distinguishable. Note the large differences for large $N$. See also Fig.~\ref{fig:numArr}. \label{tb:numArr}}
\begin{ruledtabular}
\begin{tabular}{rrr}
$N$ & $p(N)$  & $B(N)$ \\
\hline
1  &  1  &      1 \\
2  &  2  &      2 \\
3  &  3  &      5 \\
4  &  5  &     15 \\
5  &  7  &     52 \\
6  & 11  &    203 \\
7  & 15  &    877 \\
8  & 22  &   4140 \\
9  & 30  &  21147 \\
10 & 42  & 115975 
\end{tabular}
\end{ruledtabular}
\end{table}

\begin{figure}
\includegraphics[width=\columnwidth]{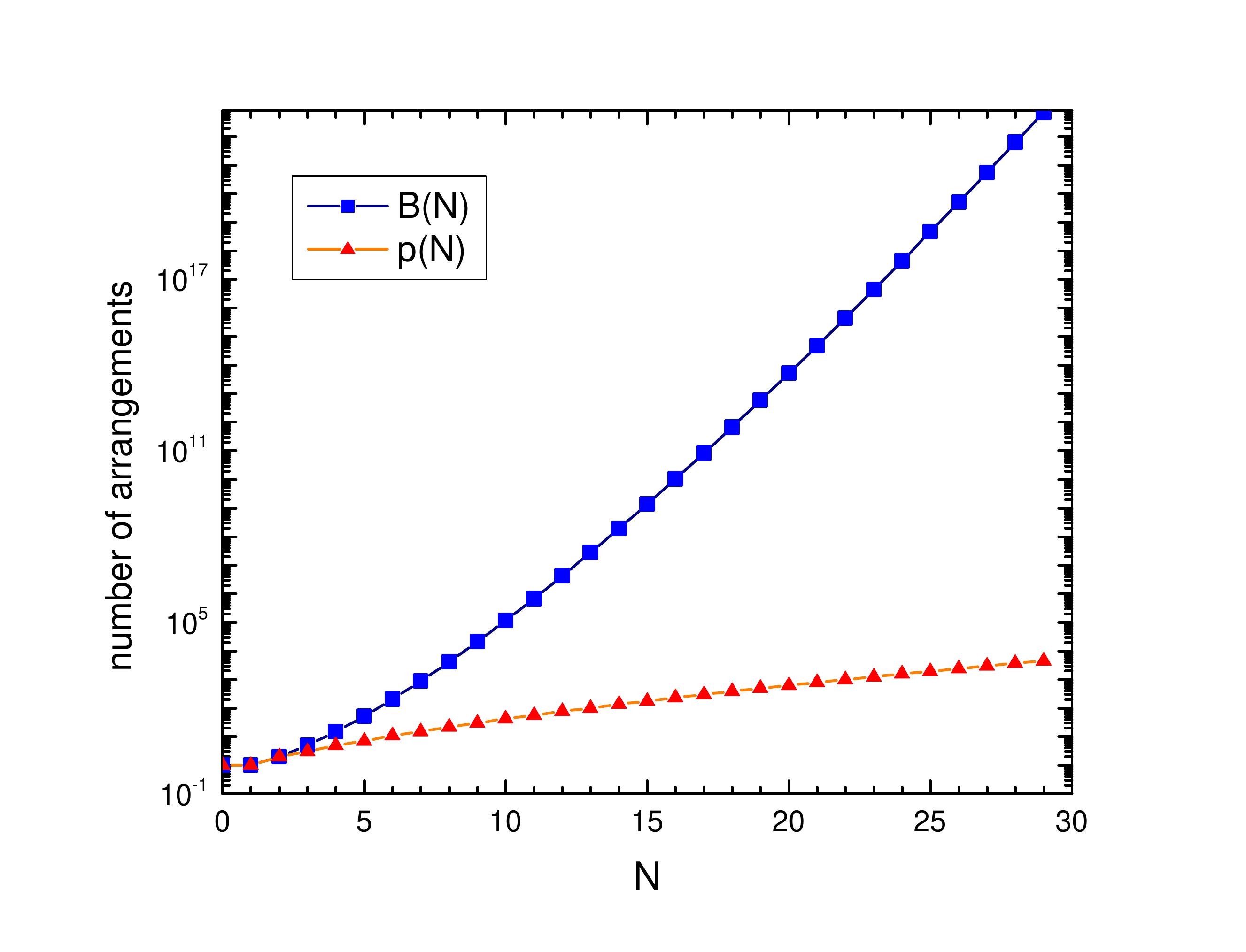}
\caption{(Color online) Number of arrangements for $N$ distinguishable particles, $B(N)$, and for $N$ identical particles, $p(N)$, assuming in both cases that the particles can bind in all configurations.  Note the log-scale for the number of arrangements and the dramatic difference between $B(N)$ and $p(N)$ for large $N$, where they have different asymptotic behaviors as described Eqs.~(\ref{eq:BNasym}) and (\ref{eq:pNasym}), respectively.
\label{fig:numArr}}
\end{figure}

The maximum number of arrangements for $N$ particles is obtained by assuming particles can bind in all configurations (groupings) and they are all distinguishable. For such $N$ distinguishable particles that can bind in all configurations, an arrangement corresponds, mathematically, to a partition of a set with $N$ distinguishable elements. The number of arrangements is given in this case by the Bell's number $B(N)$ \cite{Olver10}. It can be determined from $B(0)=1$ and through a recurrence relation \cite{Olver10}
\begin{equation}
\mathop{B\/}\nolimits\!\left(N+1\right)=\sum_{k=0}^{N}\binom{N}{k}\mathop{B\/}%
\nolimits\!\left(k\right).
\end{equation}
Examples of $B(N)$ for small $N$ are given in Table~\ref{tb:numArr}. For large $N$, it has an asymptotic behavior given by \cite{Olver10}
\begin{equation}
\mathop{B\/}\nolimits\!\left(N\right)=\frac{K^{N}e^{K-N-1}}{(1+\mathop{\ln\/}%
\nolimits K)^{1/2}}\left(1+\mathop{O\/}\nolimits\left(\frac{(\mathop{\ln\/}%
\nolimits N)^{1/2}}{N^{1/2}}\right)\right),
\end{equation}
where $K\ln K=N$ or equivalently $K=e^{\mathrm{Wp}(N)}$ where $\mathrm{Wp}(N)$ is the Lambert function \cite{Olver10}. To the lowest order, this asymptotic behavior means that $B(N)$ grows for large $N$ as $N^N$ or equivalently (see also \cite{deBruijn81})
\begin{equation}
B(N) \overset{N\to\infty}{\sim} e^{N\ln N} \;.
\label{eq:BNasym}
\end{equation}
The Bell's number $B(N)$ provides the upper limit for the number of arrangements for an $N$-body quantum system. If some of the particles only repel each other, some of the configurations (partitions of the set) would not bind, and the number of arrangements will be reduced. If some of the particles are identical, some of the configurations which would have been different for distinguishable particles will be equivalent, and the number is again reduced. In other words, there is a general constraint on the number of arrangements $\mathscr{M}(N)$,
\begin{equation}
1 \le \mathscr{M}(N) \le B(N) \;.
\label{eq:scrMconstraint1}
\end{equation}

When some or all of the particles are identical, the reduction of the number of arrangements can be dramatic. This dramatic reduction is best illustrated by considering the special case of $N$ identical particles that can bind in all configurations, which is equivalent, in this case, to being able to bind in any number. For such $N$ identical particles that can bind in any number, an arrangement corresponds mathematically to a partition of the natural number $N$ \cite{Olver10}. The number of arrangements is given in this case by the function $p(N)$: the number of partitions of the natural number $N$ \cite{Olver10}. It can be calculated from $p(0)=1$ and through a recurrence relation \cite{Olver10}
\begin{equation}
p\left(N\right)=\sum_{k=1}^{\infty}(-1)^{k+1}\left[p\left(N-\omega(k)\right)+p%
\left(N-\omega(-k)\right)\right] \;,
\end{equation}
where $\omega(k)$ are the pentagonal numbers defined by
\[
\omega(\pm k) = (3k^2\mp k)/2 \;,
\]
and $p(k)=0$ for $k<0$. Examples of $p(N)$ for small $N$ are tabulated in Table~\ref{tb:numArr}. For large $N$, $p(N)$ has a famous asymptotic behavior of Hardy and Ramanujan \cite{HR1918,Olver10}
\begin{equation}
p(N) \overset{N\to\infty}{\sim} \frac{1}{(4\sqrt{3}) N} e^{\sqrt{2/3}\pi\sqrt{N}} \;.
\label{eq:pNasym}
\end{equation}
It says that the $p(N)$ grows like $e^{\alpha\sqrt{N}}$, drastically smaller than and different from the behavior of $e^{N\ln N}$ for $B(N)$. This is illustrated in Fig.~\ref{fig:numArr}.

The combination of $p(N)$ and $B(N)$ provides a tighter constraint for $N$ particles, some distinguishable and some identical, that can bind in all configurations,
\begin{equation}
p(N) \le \mathscr{M}(N) \le B(N) \;.
\end{equation}
It is a useful constraint for most quantum systems made of $N$ neutral atoms, which, with very few exceptions involving $^3$He, are known to bind in all configurations. The wide range for $\mathscr{M}(N)$ illustrates the wide range of complexities of $N$-body quantum systems and the drastic difference between systems of distinguishable particles and identical particles.

\section{Conclusions}
\label{sec:conc}

In conclusion, we have shown that an $N$-body system with binding is qualitatively more complex and different from a system without binding. This qualitative difference necessitates the emergent concept of arrangement for its full description, specifically of the structure of its energy spectrum.

We have shown that the arrangement concept can be extrapolated to be a general concept for any $N$-body quantum system, and each arrangement can be assigned a natural number describing the energy ordering of the arrangements. We have also shown that the total number of arrangements for an $N$-body quantum system, which gives a measure of the chemical complexity of the system, depends critically on whether the particles are distinguishable or identical, with vastly different asymptotic behaviors for large $N$.

With the incorporation of the concept of arrangement, the theory of $N$-body quantum system is complete at least in principle. It allows one to rigorously formulate all states and all processes in an $N$-body quantum system. Combining it with our recent work on the concept of rigidity and its implications on $N$-body wave functions through cusp conditions \cite{Gao22a}, we have a reconstructed foundation to build quantum theories that better treat the complexities of $N$-body systems, especially those with binding, and to further probe the many fundamental questions associated with them. On the side of practical applications, the questions include the dynamics of phase transitions, such as the dynamics of condensation or evaporation \cite{LandauSTM1}. They also include more general theories of reactions especially those beyond bimolecular reactions, and better theories for catalysis. Identifying the location of an initial state in an $N$-body energy spectrum, and the arrangements and states to which the initial state is most strongly coupled is a first and necessary step towards understanding the dynamics of an $N$-body quantum system. Within a broader structure of our understanding of nature, this new foundation for $N$-body theory serves to better unify chemistry and physics. It asserts that the essence of chemistry is, at a fundamental level, about the properties of $N$-body quantum systems with binding, and can be understood within such a framework. At a deeper conceptual level, a better $N$-body quantum theory is closely related to the foundation of statistical mechanics \cite{LandauSTM1,Huang87}, complexity and the arrow of time \cite{LDR13}, the quantum-classical correspondence \cite{Zurek91,Gao99b,Schlosshauer05}, as well as the relation and the interplay between emergence and reduction \cite{Anderson72,Laughlin00}. The impact of a better $N$-body quantum theory on these questions and issues will be gradually felt and realized in future works.

\begin{acknowledgments}
This work was supported by NSF under Grant No.~PHY-1912489. 
\end{acknowledgments}

\end{document}